\documentstyle[12pt,psfig]{article}
\nopagebreak
\newcommand\etal{{{\it et al.}\/}}

\def\oneh{{\textstyle {1\over 2}}}

\def\simle{{\ \lower2pt\hbox{$\sim $}\mkern-13mu \raise2pt \hbox{$<$}\ }}

\def\eq{\begin{equation}}
\def\ee{\end{equation}}

\def\eqa{\begin{eqnarray}}
\def\eea{\end{eqnarray}}

\def\eep{(e,e$'$p)\ }
\def\een{(e,e$'$n)\ }

\begin{document}


\centerline{\Large{\bf Charge-exchange and multi-scattering effects}}

\vskip 0.5cm
\centerline{\Large{\bf in \een knockout}}

\vskip 1.5cm

\centerline{\large{P.~Demetriou, S.~Boffi, C.~Giusti and F.~D.~Pacati}}

\vskip 1.0cm

\centerline{\small Dipartimento di Fisica Nucleare e Teorica, Universit\`a
di Pavia, and}

\centerline{\small Istituto Nazionale di Fisica Nucleare, 
Sezione di Pavia, Pavia, Italy}

\vskip 1.5cm


\begin{abstract}
\noindent
Final-state interactions in \een knockout reactions in the quasi-free region are studied
by considering the multistep direct scattering of the ejectile nucleon. Primary and
multiple particle emission are included within the same model and are found to become
important with increasing excitation energy. Charge-exchange effects taken into 
account through the two-step \eep~(p,n) and three-step \eep~(p,N)~(N,n)
 processes are also found to increase with energy. A comparison with the 
results obtained with an isospin-dependent optical potential at small
excitation energies is presented. 
\end{abstract}

\bigskip




\section{Introduction}
A large number of experiments on \eep reactions have been performed 
over the past years for a wide range of nuclei and kinematical regions. In the 
quasi-free (QF) region, the experimental cross sections are well described in the
distorted-wave-impulse-approximation (DWIA), where the virtual  photon 
interacts only with the emitted nucleon while the other nucleons are spectators
and contribute to FSI only. Extensive comparisons between theoretical and
experimental cross sections, including the effects of FSI, 
have been able to provide insight on the s.p. properties of the nucleus~\cite{book96}.

On the other hand, additional and complementary information is, in principle, available from
the \een reaction. A comparison between \eep and \een data taken in similar kinematics would
allow us to study the different behaviour of protons and neutrons in nuclei. Experiments on the \een reaction have not been made so far
due to the difficulties in performing high-resolution neutron detection,
particularly since theoretical predictions have shown the \een cross sections
to be significantly smaller than the \eep cross sections. Such an experimental comparison
would however be of great interest for clarifying the reaction mechanisms, also in view of the
fact that ($\gamma, p$) and ($\gamma, n$) cross sections have been observed to be of the same
size~\cite{book96}. From the theoretical point of view it is important to investigate the relevance of those
processes which could play a different role in proton and neutron emission, and which might
hamper the interpretation of data. In a series of papers~\cite{gp89,ry89,jes94} the effects
of charge-exchange FSI was addressed. In~\cite{gp89} the \een reaction
was described as  a direct mechanism accompanied
 by a two-step process, where a proton interacts with the virtual photon
  and then undergoes a (p,n) reaction. The charge-exchange process was treated
   as an isospin-flip in the final state interaction by an isospin
 dependent optical potential. The resulting cross sections showed a small effect 
 arising from the charge-exchange contribution which decreased with
the outgoing neutron energy. This was found  to be in agreement with the results
obtained within a self-consistent Hartree-Fock and continuum RPA 
model~\cite{ry89}. A different approach using a coupled-channels method
 in the lowest missing energy range~\cite{jes94} also predicted
  small contributions from charge-exchange processes.

In this paper we aim at investigating the \een reaction in the QF region 
with particular interest in the effects of charge-exchange in FSI. The latter are described as
a series of two-body NN interactions by means of the quantum-mechanical
multistep direct theory (MSD) of ~\cite{fkk80}. Apart from 
direct neutron knockout, other processes  contribute to the neutron emission spectra: 
the neutron after having absorbed the virtual photon undergoes several multistep scatterings
before being emitted, and/or the virtual photon is absorbed by a proton which 
undergoes a charge-exchange reaction leading to neutron emission. The charge-exchange
reaction can occur immediately after the photon absorption or after some 
re-scatterings. 

At excitation energies above 50 MeV multi-nucleon emission can 
also occur through a multi-scattering mechanism. In this case the initially 
excited nucleon interacts with another nucleon exciting it to the continuum. Thus 
a secondary nucleon is emitted and
may be detected giving rise to a different energy spectrum compared to the primary
nucleon. The contribution of secondary nucleon emission to  nucleon 
energy spectra
 has been
investigated in nucleon-induced reactions by extending the theory of~\cite{fkk80} 
and has been found to be important at high excitation energies~\cite{chad94}.
 We use the same method proposed by~\cite{chad94} to estimate the contribution of
 secondary neutrons to the \een energy spectrum.  

The theory of direct nucleon knockout reactions, primary MSD and multiple MSD
reactions is presented in sect.~2. The results are given in sect.~3 and some
conclusions are discussed in sect.~4.
\section{Theory}
\subsection{Electroinduced direct nucleon knockout}
The exclusive cross section for the \een and \eep direct knockout reaction  is obtained 
in the one-photon-exchange approximation. For an ejectile electron energy $E_{k'}$ and angle
$\Omega_{k'}$ and ejectile neutron of energy $E$ and angle $\Omega$ it can be written
in terms of four structure functions~\cite{book96}
\eqa
{{\rm d}^3\sigma\over {\rm d}\Omega_{k'}{\rm d}E_{k'}{\rm d}\Omega}
&=& { 2\pi^2\alpha \over \vert{\vec q}\vert}\, \Gamma_{\rm V}\, 
K  \{ W_{\rm T} +   \epsilon_{\rm L}\, W_{\rm L} \nonumber \\
& & \nonumber \\
& & + \sqrt{\epsilon_{\rm L}(1+  \epsilon)} W_{\rm TL} \cos\phi +  
\epsilon\, W_{\rm TT}\cos  2\phi\},
\label{eq:nonfact} 
\eea
where $\Gamma_{\rm V}$ is the flux of virtual photons, $\phi$ the out-of-plane
angle of the nucleon with respect to the electron scattering plane,
\eq 
\epsilon = 
\left [ 1 +   2{\vert{\vec q}\vert^2\over Q^2} \tan^2\oneh\theta
\right]^{-1},\qquad
\epsilon_{\rm\scriptstyle L} = {Q^2\over\vert{\vec q}\vert^2} \epsilon ,
\ee
and $Q^2 = \vert{\vec q}\vert^2 - \omega^2$ is the negative mass squared of the
virtual photon defined in terms of the momentum ${\vec q}$ and energy $\omega$
transferred by the incident electron through a scattering angle $\theta$. Transitions to
discrete final states are  calculated with eq.(1)
in DWIA and are extended to the continuum by including an energy
  distribution taken from~\cite{kra90} and described in~\cite{npa}. 

\subsection{Primary multistep emission}

The details of the formalism have been presented in~\cite{npa} so in the following we
shall only give
a brief account of the formulae used.
The cross section for primary MSD emission is written as an
incoherent sum of a direct neutron knockout \een and multistep neutron
emission cross sections
\eq
\frac{{\rm d}^{4}\sigma}{{\rm d}\Omega_{k'} {\rm d}E_{k'} {\rm d}\Omega {\rm d}E} 
= 
\frac{{\rm d}^{4}\sigma^{(1)}}{ {\rm d}\Omega_{k'} {\rm d}E_{k'} {\rm d}\Omega
{\rm d}E } 
+ \sum_{n=2}^{\infty}\frac{{\rm d}^{4}\sigma^{(n)}}{ {\rm d}\Omega_{k'}
{\rm d}E_{k'} {\rm d}\Omega {\rm d}E}\,,
\label{eq:msdeep}
\ee
where the multistep ($n$-step) cross section is given
 by the convolution integral
 \begin{eqnarray}
\frac{{\rm d}^{4}\sigma^{(n)}}{ 
{\rm d}\Omega_{k'}{\rm d}E_{k'} {\rm d}\Omega {\rm d}E} 
& =&  \left(\frac{m}{4\pi^{2}}\right)^{n-1} 
\int {\rm d}\Omega_{n-1}\int {\rm d}E_{n-1}E_{n-1}\dots \nonumber \\
& & \times \int {\rm d}\Omega_{1}\int {\rm d}E_{1}E_{1} 
\frac{{\rm d}^{2}\sigma^{(1)}}{{\rm d}\Omega dE}({\rm n}, {\rm N}^{(n-1)})\dots \nonumber \\
& &\times \frac{{\rm d}^{2}\sigma^{(1)}}{{\rm d}\Omega_{2}
 {\rm d}E_{2}}({\rm N}^{(2)},{\rm N}^{(1)}) 
\frac{{\rm d}^{4}\sigma^{(1)}}{{\rm d}\Omega_{k'}{\rm d}E_{k'} {\rm d}\Omega_{1} dE_{1}}
({\rm e,e'N}^{(1)})\,,
\label{eq:msdneep}\\ \nonumber
\end{eqnarray} 
over all intermediate
energies $E_{1},\, E_{2}\dots$ and angles $\Omega_{1},\,\Omega_{2}\dots$ obeying
energy and momentum conservation rules; $m$ is the nucleon mass and N = n or p the particle
excited in the intermediate one-step reactions (p = proton and n = neutron).

\noindent
The one-step MSD cross section ${\rm d}^{2}\sigma^{(1)}/{\rm d}\Omega dE({\rm N}^{(n)}
,{\rm N}^{(n-1)})$ is calculated by extending DWBA to the continuum and is given by 

\begin{eqnarray}
\frac{{\rm d}^{2}\sigma^{(1)}}{{\rm d}\Omega {\rm
d}E} ({\rm N}^{(n)},{\rm N}^{(n-1)})= \sum_{J}(2J+1)\rho_{1p1h,J}(U)
\left\langle\frac{{\rm d}\sigma}
{{\rm d}\Omega}\right\rangle^{\rm DWBA}_{J},
\label{eq:onestep}
\end{eqnarray} 
where $J$ is the orbital angular momentum transfer, $\langle {\rm d}\sigma/
{\rm d}\Omega\rangle^{\rm DWBA}_{J}$ is the average of DWBA cross sections exciting
$1p1h$ states consistent with energy, angular momentum and parity conservation and
$\rho_{1p1h,J}(U)$ is the density of such states with residual nucleus
energy $U=E_{n-1}-E_{n}$. All possible inelastic or charge-exchange processes corresponding
 to (${\rm N}^{(n)},{\rm N}^{(n-1)}$) that can occur at the $n$th-step of FSI are estimated using eq.(5).

\subsection{Multiple emission}

Multiple pre-equilibrium emission in nucleon-induced reactions  has been addressed 
by~\cite{chad94} using the theory of Feshbach \etal~\cite{fkk80}. In their approach there
exist two types of multiple emission: ``type I'', in which more than one excited
particle is emitted immediately after a single intranuclear collision occurs; and
``type II'', where a particle is emitted after which a number of damping
transitions occur, and then a second particle is emitted and so on. Their 
calculations showed
that type II processes are relatively small compared to processes of type I and can
therefore be omitted. In the following we apply the same formalism to multiple nucleon
emission induced by electron scattering off the nucleus. We treat the multistep
scatterings and multiple emission processes exactly as in nucleon-induced reactions
assuming that the only difference in this case is the electromagnetic probe which
appears in the cross sections for primary emission (eqs.3-5).
We also restrict ourselves to processes in which
up to two particles are emitted, though the formalism can be generalized to include
more emissions.

According to~\cite{chad94}, to determine the cross section for emission of a secondary particle at an energy $E$ one starts with the cross section for producing
particle-hole ($p-h$) states at energy $U$ after primary emission at stage $n$.
The 
cross section at each $n$ is given by eq.(4). Then one determines the
probability that among such states there exists a nucleon with energy $E+E_{b.s.}$
 (where $E_{b.s.}$ is the separation energy) which can escape with transmission-coefficient probability. A basic assumption here is that all possible $p-h$
 configurations are equiprobable. The angle-integrated cross section is then given
 as the sum of contributions from each primary emission step $n$ as follows
\eqa
 \frac{{\rm d}^{3}\sigma_{mul}^{(j)}}{{\rm d}\Omega_{k'}{\rm d}E_{k'}{\rm d}E} =
  \sum_{n} \frac{{\rm d}^{3}\sigma_{mul}^{(n,j)}}{{\rm d}\Omega_{k'}{\rm d}E_{k'}
  {\rm d}E}\,,
 \eea
 where
 \eqa
 \lefteqn{\frac{{\rm d}^{3}\sigma_{mul}^{(n,j)}}{{\rm d}\Omega_{k'}{\rm d}E_{k'}
 {\rm d}E} =  \sum_{i={\rm p,n}}
\int_{U=E+E_{b.s.}}^{U_{max}} \frac{{\rm d}^{3}\sigma^{(n,i)}}{{\rm d}\Omega_{k'}{\rm d}E_{k'}{\rm d}U}} \nonumber \\
& &\times\, \left [
\frac{1}{p}\frac{\omega(1p,0,E+E_{b.s.})\omega(p-1,h,U-E-E_{b.s.})}{\omega(p,h,U)}
R^{i,j}_{n}\right ] \,T_{j}(E)\, {\rm d}U,
\eea
where $E$ is the emission energy, $i$ labels the primary particle emitted (p =
proton, n = neutron) and $j$ the multiple particle emitted. The quantity in the
square brackets is the probability of finding a particle $j$ at an energy $E+E_{b.s.}$
inside a $p-h$ configuration of energy $U$, with $\omega(p,h,U)$ being the Fermi gas
level density at excitation energy $U$. $R_{n}^{i,j}$ is the probability of
finding a nucleon of type $j$ in the $p-h$ configuration
 after primary emission of nucleon type $i$ at step $n$ and is calculated as prescribed
 by~\cite{blan}, ${\rm d}^{3}\sigma^{(n,i)}/{\rm d}\Omega_{k'}{\rm d}E_{k'}{\rm d}U$ is 
 the differential cross section of $p-h$ states after primary emission
  of a nucleon type $i$ at stage $n$. It is given as a function of the residual nucleus energy and is
  obtained by angle-integration of the cross sections of eq.(4). $T_{j}$ is a Gamow penetrability factor
  and
  describes the probability that the continuum particle $j$ escapes with an energy
  $E$.
  
The particles emitted through multiple and primary emission are given the same 
angular distribution through the following equation 
\eqa
\frac{{\rm d}^{4}\sigma^{(n,j)}_{mul}}{{\rm d}\Omega_{k'}{\rm d}E_{k'}{\rm d}\Omega {\rm
d}E} = \frac{{\rm d}^{3}\sigma_{mul}^{(n,j)}}{{\rm d}\Omega_{k'}{\rm d}E_{k'}
{\rm d}E}G(\Omega) ,
\eea
where the angular kernel $G(\Omega)$ is determined from eq.(4) as:
\eqa 
G(\Omega) = 
({\rm d}^{4}\sigma^{(n,j)}/{\rm d}\Omega_{k'}{\rm d}E_{k'}{\rm d}\Omega{\rm d}E)/
({\rm d}^{3}\sigma^{(n,j)}/{\rm d}\Omega_{k'}{\rm d}E_{k'}{\rm d}E).
\eea
 
\section{Results}
The calculations of the (e,e$'$N) direct knockout cross sections and primary 
MSD cross sections were performed using the same input parameters as in \cite{npa}. 
The distorted waves were obtained from the optical potential of~\cite{gian76} and the 
b.s. wavefunctions from a Woods-Saxon potential with the geometrical parameters
of~\cite{mou76}. The energy level densities involved in the primary MSD and multiple
emission cross sections were obtained from an equidistant Fermi-gas model with finite
hole-depth restrictions~\cite{wil71} and an average
single-particle density $g=A/13$ and the spin distribution was taken to be Gaussian with a spin
cut-off parameter $\sigma^{2}_{n}$ of~\cite{fu86}. The multiple MSD cross-sections were
calculated using computer subroutines developed by~\cite{chad97}. 

The method was applied to the \een reaction on $^{40}$Ca which is a suitable nucleus
for the statistical assumptions of the MSD theory. Identical kinematic conditions
 as those in~\cite{npa} were used, i.e, incident electron energy $E_k =  497$
MeV and electron scattering angle $\theta = 52.9^{\circ}$. We fixed the scattered electron energy
at $E_{k'} =  350$ MeV
and  worked at constant $({\vec q}, \omega)$ by varying the neutron energy
$E$ accordingly. 

In figure 1 we show the theoretical direct \een knockout and multistep 
emission angular distributions at four excitation energies. The angle $\gamma$ corresponds to the 
angle between the outgoing neutron $\vec p$ and momentum transfer $\vec q$. The 
multistep emission curves include contributions from all possible multistep 
scatterings of a n or p following photon absorption that end up
 in a neutron being emitted. For the three-step emission for example,
we take into account the (e,e$'$n$'$)~(n$'$,N)~(N,n) and the 
\eep(p,N)~(N,n) processes. At these excitation
energies we compare only the direct knockout cross sections with the primary MSD emission
cross sections.

The results show that the direct knock-out \een process is 
dominant only at the lower excitation energies (lower missing energies) and even 
then only at forward angles (missing momentum $p_{m}\leq$~200~MeV/c).
The multistep scattering cross sections increase with energy and
 scattering angles and at the higher excitation energies account for
 almost all the emission cross section. This is due to the large contributions
 from the two-step \eep(p,n) and three-step \eep(p,N)~(N,n)
 processes which involve  charge-exchange reactions. In fact, the contributions
 of these multistep processes are much larger than those from the 
 (e,e$'$n$'$)~(n$'$,n) and (e,e$'$n$'$)~(n$'$,N)~(N,n) processes.
 At the lowest missing
 energy the effect of charge-exchange contributions amounts to only 
 $\approx$~30~$\%$ of the total cross section, however it increases rapidly
 at large scattering angles and excitation energies.

It is worth comparing these results with those obtained using an isospin-dependent
optical potential to account for charge-exchange contributions as reported 
in~\cite{gp89}. 
In figure 2 we present the angular distribution for the direct knockout of a neutron
from the 1s$_{1/2}$ orbit in $^{40}$Ca using the same kinematic conditions
 as in the case of figure 1.
 The residual nucleus is left at an 
excitation energy $U$ = 22 MeV which roughly corresponds to the lowest excitation energy
included in figure 1 so a comparison is possible. The cross sections with 
charge-exchange effects from an optical potential 
 are identical
 to those obtained without charge-exchange and this can be clearly seen by 
 plotting the difference
 between the two results (dotted line) in the same figure. The effect is of the
  order of $\approx 1-2\,\%$
 in contrast with the $\approx 30\%$ effect given by multi-scattering  contributions at the same excitation energy in figure 1. Therefore with the present approach
  of explicitly including multi-scattering effects in FSI, 
 charge-exchange
 contributions are small at low missing energies confirming the dominance of
  the direct knockout mechanism in these energy regions, yet they are not as
   negligible as found in previous works.
   
 In figures 3 and 4 we show the excitation energy spectra for neutron emission 
 at two
 different sets of neutron scattering angles $\gamma = 0^{\circ},\,\phi = 0^{\circ}$ and
$\gamma = 30^{\circ},\,\phi = 0^{\circ}$ respectively. Contributions from primary MSD and 
secondary MSD emission are included.  At energies above the two-nucleon emission threshold a secondary neutron can
  be emitted along with a primary neutron or
  proton. Furthermore, primary neutron or proton
  emission result from either a \een or \eep reaction so all processes such as
  (e,e$'$n$'$)~(n$'$,xn), \eep(p,xn) including further re-scatterings are taken into
  account. At $\gamma = 0^{\circ}$ and at low excitation 
 energies
 direct neutron knockout is dominant in agreement with figure 1. However 
 primary MSD emission is also important and becomes the dominant
 process with increasing energy and scattering angle as can be seen in figure 4. Charge-exchange processes
 which give the major contribution to primary MSD emission have already 
 been included in the curves.
 Multiple particle emission also tends to increase with excitation energy and 
 scattering angles as is seen by comparing figures 3 and 4. It seems that at
 larger
 scattering angles the strength is shifted from primary MSD to multiple MSD
 processes. 
 The most important
  multiple emission contribution comes from two-step \eep~(p,xn) which is 
  comparable in magnitude with that of the primary three-and four-step MSD 
   processes even at the lower excitation energies.
  Two-step secondary neutron emission is equivalent to a knockout mechanism in which a nucleon excited by a
  one-photon exchange mechanism strikes a bound neutron and both particles are 
  emitted. In this respect it is quite similar to direct two-nucleon knockout
  which arises from  
  meson-exhange mechanisms (MEC) and short-range correlations (SRC) and has 
  also been 
  found to be important above the two-nucleon
  emission threshold for the $^{12}$C\eep reaction~\cite{gp94a}.
  
\section{Conclusions}

We have treated FSI in the $^{40}$Ca\een reaction using the MSD theory
of~\cite{fkk80}. With this approach we are able to describe cross sections in the
continuum thus our conclusions apply only to this region. We 
have shown that processes arising from an \eep reaction followed by
 charge-exchange
MSD reactions (p,n) give the main contribution over the whole energy
and angle range. Even at the lowest excitation energies and at the forward-scattering angles 
we find effects up to $30\%$ which is larger than the estimates obtained with an isospin
dependent optical potential.
At excitation energies above the two-nucleon emission threshold secondary 
neutron emission 
has also been described within a multiple MSD model. The contributing cross sections 
are comparable in magnitude with those of three- and four-step primary neutron 
emission.
 
 Charge-exchange processes in FSI have also been considered
 in two-nucleon knockout reactions. There the emission of a 
 proton-proton pair is expected to be much lower than the emission of a 
 proton-neutron pair. It has however been suggested that the former could be enhanced by
 contributions from two-step processes where a proton-neutron pair is first excited and then
 two protons are emitted via a charge-exchange process (e,e$'$pn)~(n,p). 
 So far such processes
  have been taken into account by means of an isospin-dependent optical 
 potential~\cite{gp94b} and have been found to be
  negligible. In view of the  present results for primary and 
  multiple MSD emission in one nucleon knockout reactions, it would be interesting to
  extend the MSD treatment of FSI to the (e,e$'$pp)
  and (e,e$'$pn) reaction
  to see to what extent charge-exchange in multistep 
  processes will affect the relevant cross sections.

\section*{Acknowledgements}

One of us (PD) would like to thank M.~B.~Chadwick for useful discussions and for
providing the subroutines for the calculation of the multiple emission cross sections. 
\clearpage


\clearpage


\centerline{\bf Figure captions}

\medskip
\noindent
Fig. 1. Differential cross section for the $^{40}$Ca\een reaction as a function of
the angle $\gamma$ between the emitted neutron momentum and the momentum transfer at four
different excitation energies $U$ of the residual nucleus. Thin solid line 
for the direct
\een process; thin dashed, dot-dashed and dotted lines for the two-, three- and 
four-step
processes arising from the \een reaction. The corresponding thick lines are for the
two-, three- and four-step processes respectively arising from the \eep reaction. The total result is given by the thick solid line.

\noindent
Fig. 2. Differential cross section for the $^{40}$Ca\een reaction as a function of
the angle $\gamma$ between the emitted neutron momentum and the momentum transfer where the ejectile neutron
is emitted from the 1s$_{1/2}$ orbit. The solid line corresponds to the direct \een process
without charge-exchange effects and the dotted line represents the difference between the
cross sections with and without charge-exchange effects obtained from an isospin
dependent optical potential.

\noindent
 Fig. 3. Excitation energy spectra for the $^{40}$Ca\een reaction at neutron 
 scattering angles $\gamma = 0^{\circ},\,\phi = 0^{\circ}$. The solid line is
for the direct \een knockout and thin dashed, dot-dashed and dotted lines for the
two-, three- and four-step primary MSD processes respectively.
 The corresponding thick lines are for the two-, three- and four-step
 multiple MSD processes respectively.

\noindent
 Fig. 4. Same as Fig. 3 but for neutron scattering angles $\gamma = 30^{\circ},\,
 \phi = 0^{\circ}$ .


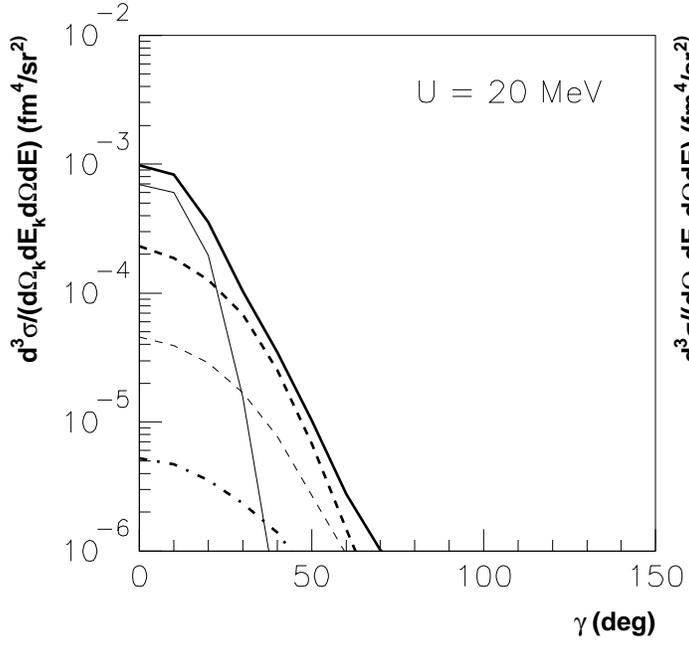
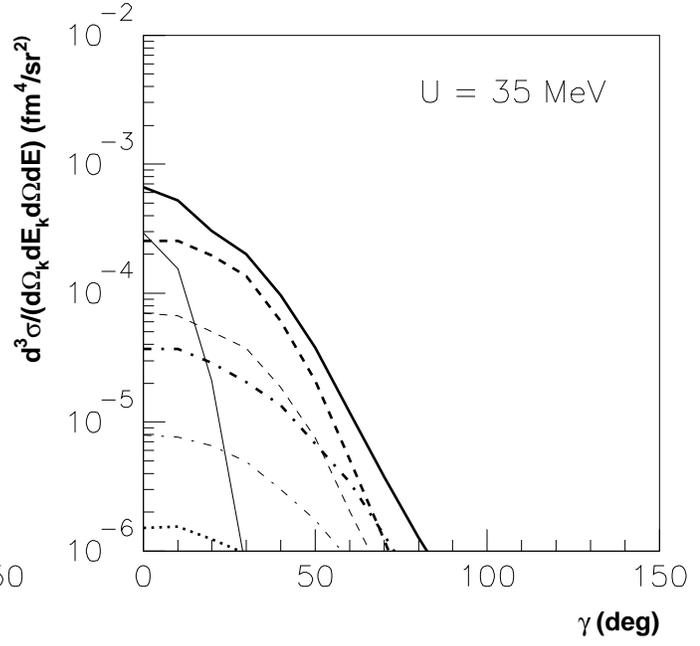
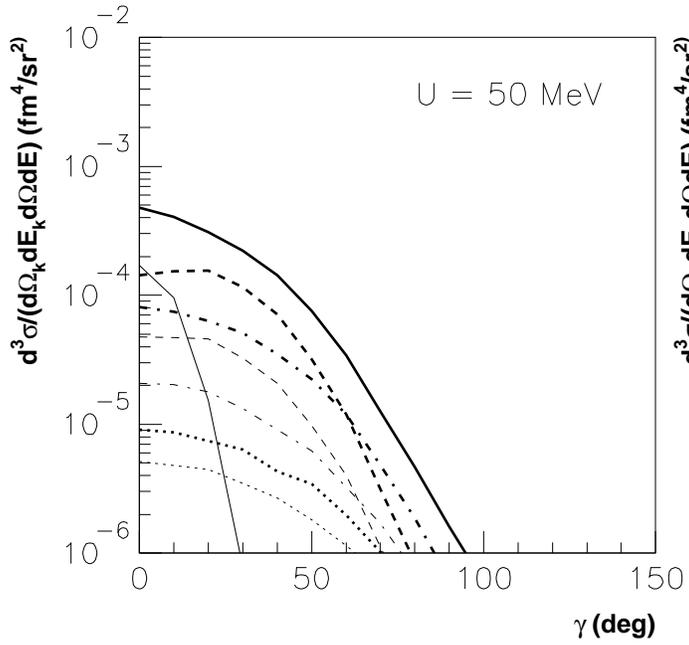
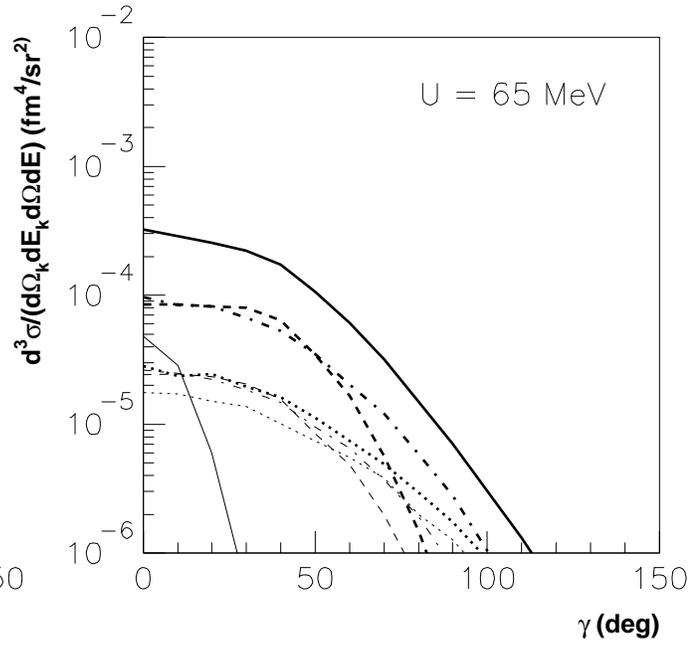
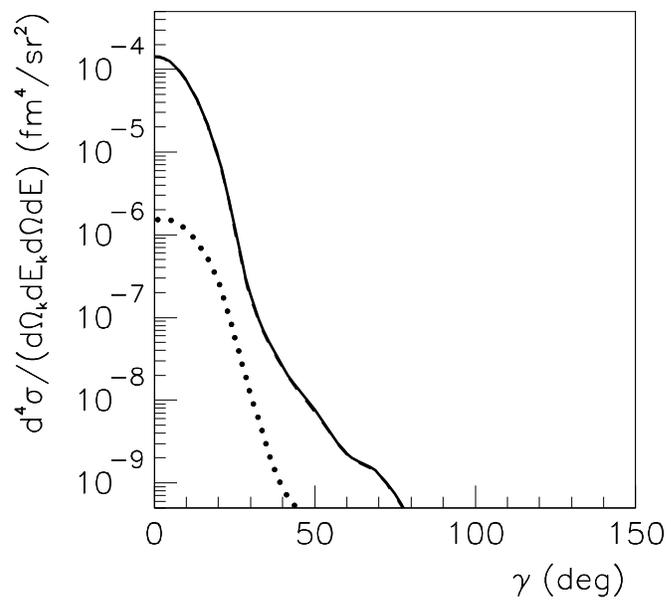
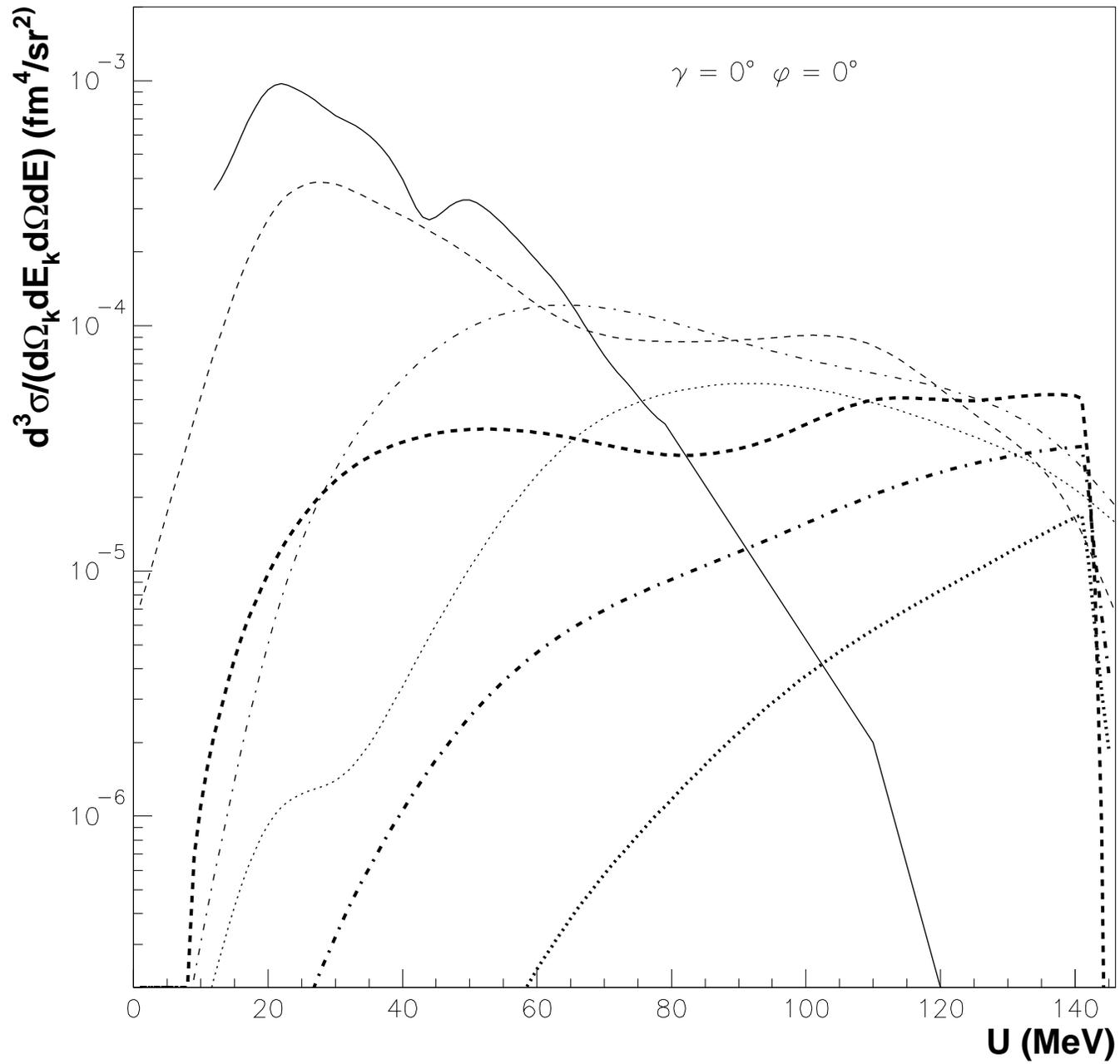
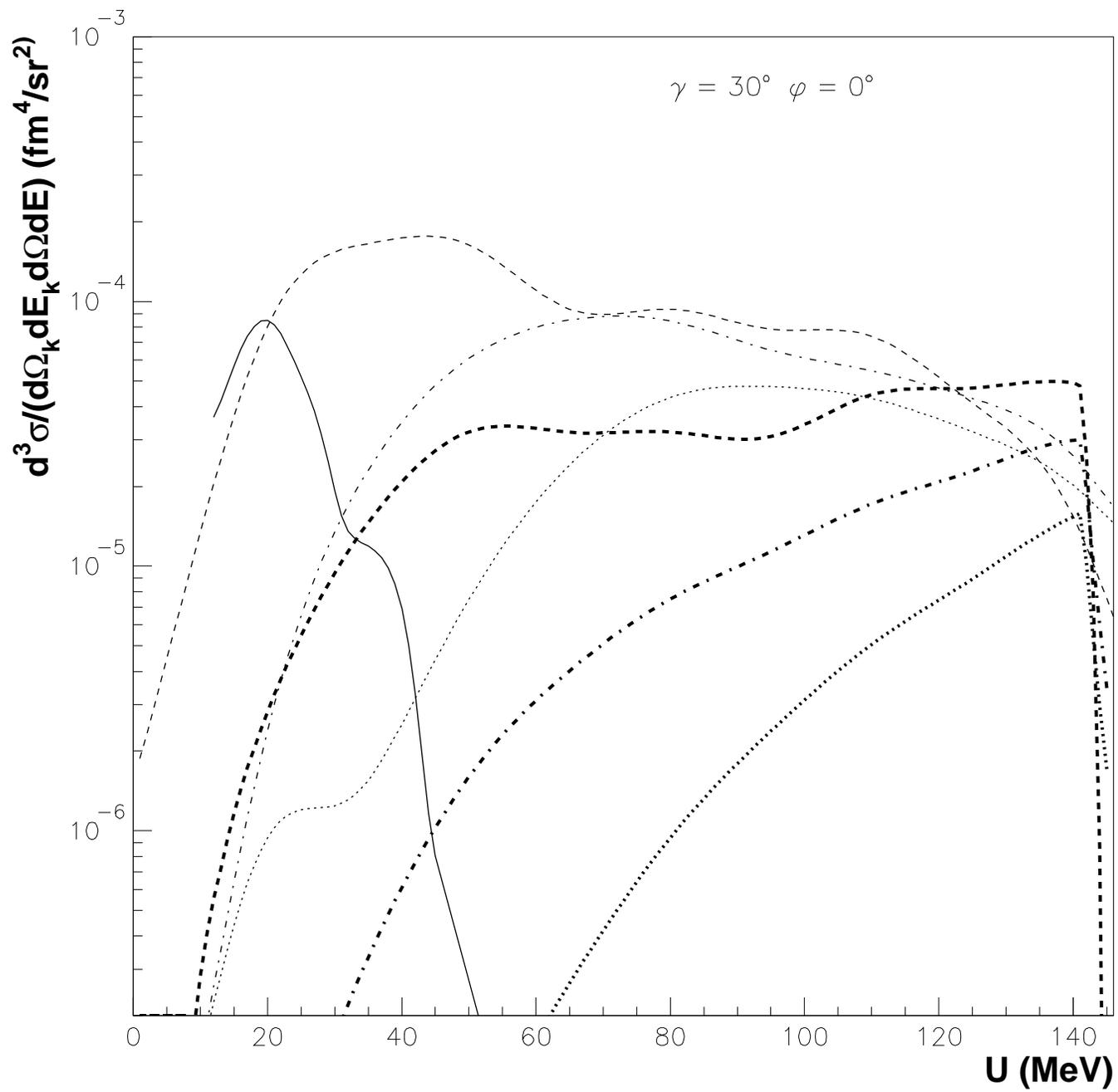
\end{document}